\documentclass[12pt]{article}

\usepackage[a4paper,text={16.8cm,22.4cm}]{geometry}
\usepackage{amsmath,amsfonts,slashed,amssymb,tikz,bm,psfrag,graphicx,color,dsfont}
\usepackage{multicol}
\usepackage{float}
\usepackage{ulem}
\usepackage{mathrsfs}
\usepackage{wick}
\usepackage{xcolor}
\usepackage{cite}
\usepackage{multirow}

\RequirePackage[sort&compress,square,comma,numbers]{natbib}
\allowdisplaybreaks
\newcommand{\nn}{\nonumber}

\begin{document}

\begin{titlepage}

\begin{flushright}
\normalsize
April 4, 2021
\end{flushright}

\begin{center}
\Large\bf
Factorization of radiative leptonic $D$-meson decay with sub-leading power corrections
\end{center}

\vspace{0.1cm}
\vspace{0.5cm}
\begin{center}
{\bf Long-Sheng Lu} \\
\vspace{0.7cm}
{ School of Physics, Nankai University, 300071 Tianjin, China}
\end{center}

\vspace{0.2cm}

\begin{abstract}
In this work, we calculate the sub-leading power contributions to the radiative leptonic $D\to\gamma \,\ell \,\nu$ decay. For the first time, we provide the analytic expressions of next-to-leading power contributions and the error estimation associated with the power expansion of ${\cal O}(\Lambda_{\rm QCD}/m_c)$. In our calculation, we adopt two different models of the $D$-meson distribution amplitudes $\phi_{D,\rm I}^+$ and $\phi_{D,\rm II}^+$. Within the framework of the QCD factorization as well as the dispersion relation, we evaluate the soft contribution up to the next-to-leading logarithmic accuracy, and the higher-twist contribution from the two-particle and three-particle distribution amplitudes is also considered. Finally, we find that all the sub-leading power contributions  are significant at $\lambda_D(\mu_0)=354\,\rm MeV$, and the next-to-leading power contributions will lead to $143\%$ in $\phi_{D,\rm I}^+$ and $120\%$ in $\phi_{D,\rm II}^+$ corrections to leading power vector form factors with $E_{\gamma}=0.5\,\rm GeV$. As the corrections from the higher-twist and local sub-leading power contributions will be enhanced with the growing of the inverse moment, it is difficult to extract an appropriate inverse moment of the $D$-meson distribution amplitude. The predicted branching fractions are $(1.88_{-0.29}^{+0.36})\times10^{-5}$ for $\phi_{D,\rm I}^+$ and $(2.31_{-0.54}^{+0.65})\times10^{-5}$ for $\phi_{D,\rm II}^+$, respectively.

\end{abstract}

\vfil

\end{titlepage}

\newpage
\section{Introduction}

The radiative leptonic heavy meson decay plays an important role in our understanding of the strong and weak interactions, and also provides the background of pure leptonic decays. It is apparent that the corresponding decay amplitudes depend on the Cabibbo-Kobayashi-Maskawa (CKM) matrix elements, the Fermi coupling constant, and the non-perturbative QCD dynamics.
Although we can extract CKM matrix elements from the pure leptonic heavy meson decay processes, it is hard to measure these processes due to the well-known helicity suppression effect. On the contrary, the radiative leptonic decays are not subject to helicity suppression due to the photon emission off the charged particles. In addition, the radiative leptonic decays are of interest on their own for exploring the factorization properties of heavy quark decays.

In the heavy quark limit, both the $B$- and $D$-meson decays could be studied within the factorization approach, the leading power (LP) predictions of $D\to e\,\nu_e\,\gamma$ are presented in \cite{Yang:2014rna,Yang:2016wtm}, where the final state photon can be either hard or soft. In 2017, the BES-III Collaboration measured the radiative leptonic $D^+\to\gamma\, e^+\,\nu_e$ decay \cite{Ablikim:2017twd}, and the upper limit on the branching fraction is about $3.0\times10^{-5}$ with $E_{\rm cut}=10 \, {\rm MeV}$. This result is in agreement with the  LP predictions \cite{Yang:2014rna,Yang:2016wtm}. However, as the charm quark mass is not much larger than $\Lambda_{\rm QCD}$, the expansion of inverse $m_c$ will work less effectively compared with $B$-meson decays. Therefore, a further study with more careful treatment of the power suppressed contribution is required. Since the energy release in $D$-meson decays is not sufficiently large, some alternative methods based on non-perturbation approach are also employed in the radiative leptonic $D$-meson decays, such as light-front quark model (LFQM), non-relativistic constituent quark model (NRQM), relativistic independent quark model (RIQM), etc. \cite{Geng:2000if,Barik:2009zza,Lu:2002mn,Atwood:1994za}. In addition, this process has also been studied using the perturbative QCD (pQCD) approach \cite{Korchemsky:1999qb}. Most of the predictions are within the upper limit of the experimental measurement.

In this work, we will study the radiative leptonic $D$-meson decay within the framework of the QCD factorization (QCDF) \cite{Collins:1981uw,Collins:1989gx,Collins:1996fb,Beneke:1999br,Beneke:2001ev} as well as the dispersion relation \cite{Khodjamirian:1999abc,Braun:2012kp,Wang:2016qii}.
In the framework of QCDF, one can separate long-distance and short-distance contributions via a simultaneous expansion in the power of strong coupling constant and $\Lambda_{\rm QCD}/m_Q$, and the factorization approach has been applied to various radiative heavy meson decays \cite{Korchemsky:1999qb,Hill:2003a,Lunghi:2003,Ma:2006a,Feldmann:2017izn,Atwood:1994za,Beneke:2001at,Beneke:2004dp,
DescotesGenon:2002mw,Bosch:2003fc,Wang:2018wfj,Shen:2020hsp}, and other processes such as factorization of the correlation function in light-cone sum rules \cite{Wang:2015vgv,Gao:2019lta,Wang:2017ijn,Lu:2018cfc,Wang:2017jow}. Since the power suppressed contribution is expected to be very important in  $D$-meson decay, we will perform a careful investigation on the ``soft" contribution, the local contribution, and the higher-twist contribution. Although the precision of the predictions with the factorization approach is limited due to the low energy scale, the inclusion of power suppressed contributions will significantly improve the reliability of the theoretical results.

Our paper is structured as follows. In section 2, we introduce the theoretical framework. Then we calculate the two-particle LP and soft contributions to next-to-leading logarithm (NLL) accuracy and evaluate the higher-twist (HT) contribution from two-particle and three-particle light-cone distribution amplitudes (LCDA). Section 3 is devoted to our numerical analysis, where we discuss the NLP contributions in detail and discuss the photon energy and inverse-moment dependence of form factors. Section 4 will be reserved for our conclusion.

\section{Formalism}

The radiative leptonic decay amplitude for the $D$-meson can be written as
\begin{eqnarray}
{\cal A}(D \to \gamma \, \ell \, \nu )
=\frac{G_F \, |V_{cd}|} {\sqrt{2}} \, \left \langle \gamma \, \ell \, \nu |
\left [ \bar{\ell} \, \gamma_{\mu} \, (1- \gamma_5) \, \nu  \right ] \,
\left [ \bar d \, \gamma^{\mu} \, (1- \gamma_5) \, c \right ] | D(p+q) \right \rangle  ,
\label{def: decay amplitude}
\end{eqnarray}
where $G_F$ is the Fermi coupling constant, and $|V_{cd}|$ is the CKM matrix element. We consider a $D$-meson with momentum $p + q$, where $p$ and $q=p_{\ell}+p_{\nu}$ denote the momenta of the photon and lepton pair, respectively. In the $D$-meson rest frame, we can decompose the four-momenta of photon and lepton pair in the light-cone coordinate
\begin{equation}\label{LCVector}
  p_{\mu}=\frac{n\cdot p}{2}\bar{n}_{\mu}\equiv E_{\gamma} \bar{n}_{\mu},\ \  q_{\mu}=\frac{n\cdot q}{2}\bar{n}_{\mu}+\frac{\bar{n}\cdot q}{2}n_{\mu},
\end{equation}
and the velocity vector of $D$-meson is $v_{\mu}=(p+q)_{\mu}/m_D=(n_{\mu}+\bar{n}_{\mu})/2$, where $n_{\mu}$ and $\bar{n}_{\mu}$ satisfy $n\cdot n=\bar{n}\cdot\bar{n}=0$ and $n\cdot\bar{n}=2$.

One could compute the $D\to\gamma \,\ell\,\nu$ amplitude to the first order of the electromagnetic interaction \cite{Wang:2016qii}
\begin{eqnarray}
{\cal A}(D \to \gamma \, \ell \ \nu)
={G_F \, |V_{cd}| \over \sqrt{2}} \, \left ( i \, g_{em} \, \epsilon_{\nu}^{\ast}  \right )
\bigg \{ T^{\nu \mu}(p, q) \, \overline \ell \, \gamma_{\mu} \, (1-\gamma_5)  \nu
+ Q_{\ell} \, f_D \,\overline \ell \, \gamma^{\nu} \, (1-\gamma_5)  \nu  \bigg \}.
\label{original B to gamma l nu amplitude}
\end{eqnarray}
Then we rewrite the hadronic tensor in terms of $F_V$ and $F_A$
\begin{equation}\label{tensornew}
\begin{split}
T_{\mu\nu}(p,q)=&-i\int d^4 x \, e^{i p \cdot x}  \,\langle 0 | {\rm T} \{j_{\mu, \rm{em}}(x),
\left [\bar d \,\gamma_{\nu} (1-\gamma_5) c \right ] (0) \} |  D(p+q) \rangle \,,\\
               =&\epsilon_{\mu \nu \rho \sigma}\,p^{\rho}\,v^{\sigma}F_V-i\big [g_{\mu\nu}\,p\cdot v-v_{\mu}\,p_{\nu} \big ]\hat{F}_A-i\frac{v_{\mu}\,v_{\nu}}{p\,v}f_D\,m_D+p_{\mu}\mbox{-} \rm terms . . .
\end{split}
\end{equation}
with $\epsilon^{0123}=+1$, where $j_{\mu,em}=\sum_qQ_q\bar{q}\,\gamma_{\mu}\,q$ is the electromagnetic current, and the last term will be canceled for the sake of $\epsilon^*\cdot p=0$. In the above equation, we have used the Ward Identity $p_{\nu}\,T^{\nu\mu}(p,q)=-(Q_c-Q_d)f_D\,p_D^{\mu}$. We can redefine
\begin{equation}\label{FAhattoFA}
F_A(n \cdot p)=\hat{F}_A(n \cdot p)+\frac{Q_{\ell} f_D}{v\cdot p},
\end{equation}
so the hadronic tensor can be written as
\begin{eqnarray}
T_{\nu \mu}(p, q) &\rightarrow& \epsilon_{\mu \nu \rho \sigma}
\, p^{\rho} \, v^{\sigma} \, F_V(n \cdot p)
-i \left [ g_{\mu \nu} \,v \cdot p - v_{\nu} \, p_{\mu} \right ] \,F_A(n\cdot p) +iQ_{\ell} \, f_D \, g_{\mu \nu} \,,
\end{eqnarray}
the last term in the above can precisely cancel the second term in Eq. (\ref{original B to gamma l nu amplitude}). At last, we can write the differential decay rate of $D\rightarrow \gamma\, \ell \,\nu$ in terms of $F_V$ and $F_A$
\begin{eqnarray}
\frac{d \, \Gamma}{ d \, E_{\rm \gamma}} \left ( D \to \gamma\, \ell \,\nu \right )
=\frac{\alpha_{em} \, G_F^2 \, |V_{cd}|^2}{6 \, \pi^2} \, m_D \, E_{\gamma}^3 \,
\left ( 1- \frac{2 \, E_{\gamma}}{m_D} \right ) \,
\left [ F_V^2(n \cdot p) + F_A^2(n \cdot p) \right ] \,.
\end{eqnarray}
The following task is to compute the form factors of the photon radiated from the down quark and charm quark.

\begin{figure}
  \centering
  \includegraphics[width=0.6\textwidth]{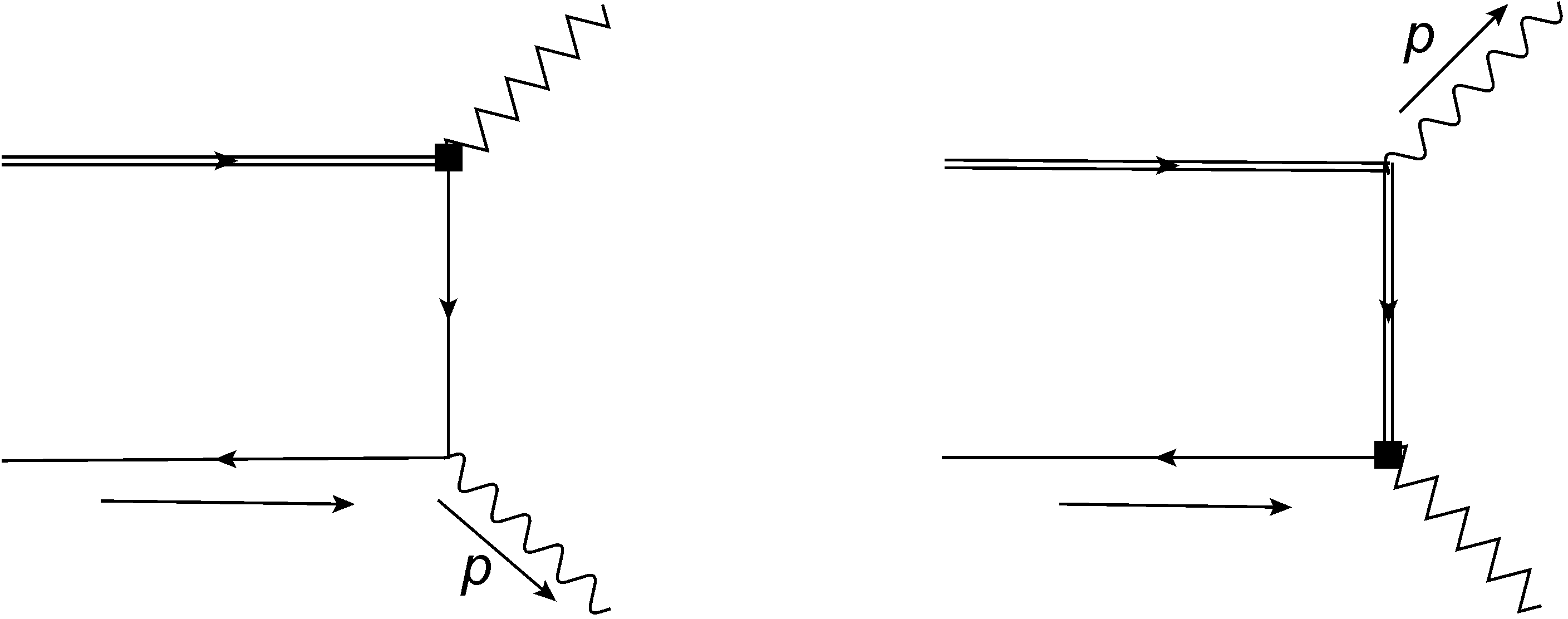}
  \caption{The Feynman diagram of the two-particle contributions at tree-level, where the double line represents the charm quark, the single line represents the down quark, and the square boxes refer to the weak vertex. }\label{treelevelfig}
\end{figure}

\subsection{Dispersion relation for the sub-leading power contribution }
Now we will evaluate the leading power and sub-leading power form factors in the framework of the dispersion relation. This approach was proposed in \cite{Khodjamirian:1999abc} for the calculation of $\gamma^*\,\gamma\to\pi$ form factors, and was applied to various processes \cite{Wang:2016qii,Beneke:2018wjp,Shen:2020hfq}. In the dispersion approach, the photon in the final state of $D\to\gamma \,\ell \,\nu$ become a space-like photon ($p^2<0$), we can treat this process perturbatively. In the framework of heavy quark effective theory (HQET), the leading power form factors of two-particle tree-level contributions can be obtained by calculating the first diagram in figure \ref{treelevelfig} with a photon radiated from the down quark
\begin{eqnarray}
&& \{F_{V}^{D \to \gamma^{\ast}}(n \cdot p, \bar n \cdot p)
,\hat{F}_{A}^{D \to \gamma^{\ast}}(n \cdot p, \bar n \cdot p)\} \nonumber \\
&&= \frac{Q_d \, F_D(\mu) \, m_D}{n \cdot p} \,
\, \int_0^{\infty} \, d \omega \, \frac{\phi_D^{+}(\omega, \mu)}{\omega - \bar n \cdot p - i 0} \,
+{\cal O}(\alpha_s, \Lambda/m_c) \,,
\label{correlator: QCD at tree level}
\end{eqnarray}
where $\phi_D^{+}(\omega, \mu)$ is the $D$-meson LCDA \cite{Grozin:1997am,Beneke:2001NPB,Beneke:2006YDS}
\begin{eqnarray}\label{D-meson LCDA}
i \, F_D(\mu) \, m_D \, \phi_D^{+}(\omega, \mu)
= {1 \over 2 \, \pi} \, \int_0^{\infty} d t \, e^{i \omega \, t } \,
\langle 0 | (\bar q_s \, Y_s)(t \, \bar n) \, \!  \not  {\bar n} \,  \gamma_5 \,
(Y_s^{\dag} \, c_v)(0) |  D(p+q) \rangle  \,.
\end{eqnarray}
In the above, $Y_s(t \, \bar n)$ is the soft Wilson line, and $F_D(\mu)$ is the HQET decay constant
\begin{eqnarray}
F_D(\mu)&=& [ K(\mu)   ]^{-1} \, f_D \,,\nonumber \\
K(\mu)&=& 1 +  {\alpha_s(\mu) \, C_F \over 4 \, \pi} \,\left [3\, \ln{m_c \over \mu} -2  \right ].
\label{matching condition for the fB}
\end{eqnarray}

The QCD spectral density in our calculation is extracted from (\ref{correlator: QCD at tree level})
\begin{eqnarray}\label{QCD spectral density}
&&{\rm Im}\, F_V^{D \to \gamma^{\ast}}(n \cdot p, \omega^{\prime}) =
{\rm Im}\,\hat{F}_A^{D \to \gamma^{\ast}}(n \cdot p, \omega^{\prime}) \nonumber \\
&&=\pi \frac{Q_u \, F_D(\mu) \, m_D}{n \cdot p} \, \phi_D^{+}(\omega^{\prime}, \mu) +{\cal O}(\alpha_s, \Lambda/m_c) \,,
\end{eqnarray}
and the expression (applicable to $F_V$ and $F_A$) of the hadronic dispersion relation is given by
\begin{eqnarray}\label{hadronic despersion relation}
F_{D\to\gamma^*}(n \cdot p, \bar n \cdot p)=\frac{f_{\rho}F_{D\to\rho}(q^2)}{m_{\rho}^2-p^2}+{1\over\pi} \int_{\omega_s}^{\infty}d\omega\frac{{\rm Im}F_{D\to\gamma^*}(n \cdot p, \omega^{\prime})}{\omega_s-\bar n \cdot p}\, .
\end{eqnarray}
According to the parton-hadron duality assumption, the spectral density ${\rm Im}\,F_{D\to\gamma^*}(n \cdot p, \bar n \cdot p)$ for $F_V$ ($F_A$) is the same as the QCD spectral density ${\rm Im}\, F_V^{D \to \gamma^{\ast}}(n \cdot p, \omega^{\prime})$ (${\rm Im}\, F_A^{D \to \gamma^{\ast}}(n \cdot p, \omega^{\prime})$) for $\omega>\omega_s$. In the above, we have combined the contributions of $\rho$ and $\omega$ because of the assumption that $m_\rho\simeq m_\omega$. Equating (\ref{correlator: QCD at tree level}) and (\ref{hadronic despersion relation}) at $\bar{n}\cdot p=0$, one obtain the relation of the $D\to\rho$ form factor $F_{D\to\rho}(q^2)$ and the QCD spectral density. Performing the Borel transformation, we have
\begin{eqnarray}
F_V(n \cdot p) &&= {1 \over \pi} \, \int_{0}^{\infty} \,\,d \omega^{\prime} \,\,  \frac{1}{\omega^{\prime}} \,
\, \left [{\rm Im}_{\omega^{\prime}} \, F_V^{D \to \gamma^{\ast}}(n \cdot p, \omega^{\prime}) \right ]  \nonumber \\
&& + {1 \over \pi} \, \int_0^{\omega_s} \,\,d \omega^{\prime} \,  \left \{ \frac{n \cdot p}{m_{\rho}^2} \,
{\rm Exp} \left [{m_{\rho}^2 - \omega^{\prime} \, n \cdot p \over n \cdot p \, \omega_M} \right ]
- {1 \over \omega^{\prime}} \right \} \,
 \left [{\rm Im}_{\omega^{\prime}} \, F_V^{D \to \gamma^{\ast}}(n \cdot p, \omega^{\prime}) \right ] ,
\hspace{0.2 cm}  \label{modified master formula of FV} \\
\hat{F}_A(n \cdot p) &&= {1 \over \pi} \, \int_{0}^{\infty} \,\,d \omega^{\prime} \,\,  \frac{1}{\omega^{\prime}} \,
\, \left [{\rm Im}_{\omega^{\prime}} \, \hat{F}_A^{D \to \gamma^{\ast}}(n \cdot p, \omega^{\prime}) \right ]  \nonumber \\
&& + {1 \over \pi} \, \int_0^{\omega_s} \,\,d \omega^{\prime} \,  \left \{ \frac{n \cdot p}{m_{\rho}^2} \,
{\rm Exp} \left [{m_{\rho}^2 - \omega^{\prime} \, n \cdot p \over n \cdot p \, \omega_M} \right ]
- {1 \over \omega^{\prime}} \right \} \,
 \left [{\rm Im}_{\omega^{\prime}} \, \hat{F}_A^{D \to \gamma^{\ast}}(n \cdot p, \omega^{\prime}) \right ] ,
\label{modified master formula of FAhat}
\end{eqnarray}
where $\omega_M$ is the Borel mass, and $\omega_s$ is the effective threshold.

Beyond tree level, the factorization formulae of the $D\to\gamma^* \,\ell\,\nu$ form factors at leading power are
\begin{eqnarray}
&& F_{V}^{D \to \gamma^{\ast}}(n \cdot p, \bar n \cdot p)
= \hat{F}_{A}^{D \to \gamma^{\ast}}(n \cdot p, \bar n \cdot p)  \nonumber  \\
&& = \frac{Q_d \, F_D(\mu) \, m_D}{n \cdot p} \, C_{\perp}(n \cdot p, \mu)  \,
\, \int_0^{\infty} \, d \omega \, \frac{\phi_D^{+}(\omega, \mu)}{\omega - \bar n \cdot p - i 0} \,
J_{\perp}(n \cdot p, \bar n \cdot p , \omega, \mu) + ... \,,
\label{master formulae for one-loop two particle factorization formula}
\end{eqnarray}
where the hard function $C_{\perp}(n \cdot p, \mu)$ and jet function $J_{\perp}(n \cdot p, \bar n \cdot p , \omega, \mu)$ are extracted simultaneously by perturbative matching with the method of regions.
The expressions of the hard function and jet function at one loop have been calculated in \cite{Wang:2016qii}
\begin{eqnarray}
 C_{\perp} &=&1- \frac{\alpha_s \, C_F}{4 \, \pi}
\bigg [ 2 \, \ln^2 {\mu \over n \cdot p} + 5 \, \ln {\mu \over m_c}
-2 \, {\rm Li}_2 \left ( 1-{1 \over r} \right )  - \ln^2  r  \nonumber \\
&&  + \,  {3 r -2 \over 1 -r}  \, \ln r + {\pi^2 \over 12} + 6  \bigg ] \,,
\label{one loop hard function}
\end{eqnarray}
\begin{eqnarray}
J_{\perp} &=& 1 + {\alpha_s \, C_F \over 4 \, \pi} \,
\bigg \{ \ln^2 { \mu^2 \over n \cdot p \,  (\omega - \bar n \cdot p)}  - {\pi^2 \over 6} - 1  \nonumber \\
&& - {\bar n \cdot p \over \omega} \, \ln {\bar n \cdot p - \omega \over \bar n \cdot p } \,
\left [ \ln { \mu^2 \over -p^2} + \ln { \mu^2 \over n \cdot p \,  (\omega - \bar n \cdot p)} + 3 \right ]  \bigg \}  \,,
\label{one loop jet function}
\end{eqnarray}
where $r=n\cdot p/m_c$. Applying the Renormalization Group (RG) approach in the momentum space, we improve the factorization formulae for the form factors to NLL accuracy
\begin{eqnarray}
&& F_{V}^{D \to \gamma^{\ast}}(n \cdot p, \bar n \cdot p)
= \hat{F}_{A}^{B \to \gamma^{\ast}}(n \cdot p, \bar n \cdot p)  \nonumber  \\
&& = \frac{Q_d \, m_D\,f_D}{n \cdot p} \, C_{\perp}(n \cdot p, \mu_{h1})K^{-1}(\mu_{h2})U(n \cdot p, \mu_{h1}, \mu_{h2}, \mu) \nonumber \\
&& \hspace{0.5 cm} \times \,\int_0^{\infty} \, d \omega \,
 \frac{\phi_D^{+}(\omega, \mu)}{\omega - \bar n \cdot p - i 0} \,
J_{\perp}(n \cdot p, \bar n \cdot p , \omega, \mu) + ... \,,
\label{resummation improved factorization formula}
\end{eqnarray}
where the factorization scale is chosen as a hard-collinear scale $\mu\sim\sqrt{\Lambda_{\rm QCD}\, m_c}$, and hard scales $\mu_{h1}$ and $\mu_{h2}$ are of order $m_c$. $U(n \cdot p, \mu_{h1}, \mu_{h2}, \mu)=U_1(n \cdot p, \mu_{h1}, \mu)\,U_2^{-1}(n \cdot p,\mu_{h2}, \mu)$ is the renormalization group equation of the hard function. The first factor $U_1(n \cdot p, \mu_{h1}, \mu)$ (Appendix \ref{b}) is given by solving
\begin{eqnarray}\label{NLL_formfactor}
\frac{d U_1(n \cdot p, \mu_{h1}, \mu)}{d \ln \mu} \,= \left [ \Gamma_{\rm {cusp}}(\alpha_s) \, \ln{\mu \over n \cdot p}
+ \gamma(\alpha_s)  \right ] \, U_1(n \cdot p, \mu_{h1}, \mu) \,
\end{eqnarray}
with $U_1(\mu,\mu)=1$ as the initial condition, where $\Gamma_{\rm {cusp}}(\alpha_s)$ and $\gamma(\alpha_s)$ are the cusp anomalous dimension and anomalous dimension, respectively. The second factor $U_2(n \cdot p,\mu_{h2}, \mu)$ is given by setting the cusp anomalous dimension in the expression of $U_1(n \cdot p, \mu_{h1}, \mu)$ to zero. The explicit expressions of the two factors could be given by replacing $E_{\gamma}$ of $U_1(n \cdot p, \mu_{h1}, \mu)$ in \cite{Beneke:2011nf} with $n\cdot p/2$.

Now we can get the dispersion relation of the $D\to\gamma$ form factors by setting $\bar{n}\cdot p$ in (\ref{NLL_formfactor}) to zero and integrating the convolution integrals in the spectral representations (Appendix \ref{appendix a})
\begin{eqnarray}
&& F_{V, 2P}(n \cdot p)
= \hat{F}_{A,2P}(n \cdot p)  \nonumber  \\
&&= \frac{Q_d \, m_D\,f_D}{n \cdot p} \, C_{\perp}(n \cdot p, \mu_{h1})K^{-1}(\mu_{h2})U(n \cdot p, \mu_{h1}, \mu_{h2}, \mu) \nonumber \\
&& \ \ \,\times \, \bigg \{   \,\int_0^{\infty} \, d \omega \,
 \frac{\phi_D^{+}(\omega, \mu)}{\omega} \,
J_{\perp}(n \cdot p, 0, \omega, \mu) \nonumber \\
&& \ \ \, + \int_0^{\omega_s} \,\,d \omega^{\prime} \,\,  \left [ \frac{n \cdot p}{m_{\rho}^2} \,
{\rm Exp} \left [{m_{\rho}^2 - \omega^{\prime} \, n \cdot p \over n \cdot p \, \omega_M} \right ]
- {1 \over \omega^{\prime}} \right ] \, \phi_{D, {\rm eff}}^{+}(\omega^{\prime},\mu) \, \bigg \}  \,, \nonumber \\
&&\equiv F_{V, \rm LP}^{\rm NLL}(n \cdot p) + F_{V, \rm{NLP}}^{\rm{soft}}(n \cdot p) \,,
\label{NLL 2-particle contribution to form factors}
\end{eqnarray}
where $F_{V, \rm LP}^{\rm NLL}(n \cdot p)$ and $F_{V, \rm{NLP}}^{\rm{soft}}(n \cdot p)$ are expressions including the first part and second part in the brace only, respectively. The convolution integral \cite{Beneke:2011nf} in LP form factor is expressed as
\begin{eqnarray}
&& \int_0^{\infty} \, d \omega \, \frac{\phi_D^{+}(\omega, \mu)}{\omega} \,  \,
J_{\perp}(n \cdot p, 0, \omega, \mu)  \nonumber \\
&& = \lambda_D^{-1}(\mu) \, \bigg \{  1 + {\alpha_s(\mu) \, C_F \over 4 \,  \pi} \,
\bigg [\sigma_2(\mu) + 2 \, \ln {\mu^2 \over n \cdot p \, \mu_0} \, \sigma_1(\mu)
+ \ln^2 {\mu^2 \over n \cdot p \, \mu_0} - {\pi^2 \over 6} -1 \bigg ] \bigg \}  \,,
\label{QCDF for B to gamma FFs}
\end{eqnarray}
where the definition of the inverse moment is
\begin{equation}\label{inversemoment}
  \frac{1}{\lambda_D(\mu)}=\int_{0}^{\infty}d\omega\frac{\phi_D^+(\omega,\mu)}{\omega}.
\end{equation}
The effective distribution amplitude $\phi_{D,\rm eff}^{+} (\omega^{\prime},\mu)$ \cite{Beneke:2018wjp} in the soft form factor reads
\begin{eqnarray}
\phi_{D,\rm eff}^{+} (\omega^{\prime},\mu)
&=& \phi_{D}^{+}(\omega^{\prime},\mu)
+ \frac{\alpha_{s}(\mu)C_{F}} {4\pi}
\bigg\{
\Big(
\ln^{2}\frac{\mu^{2}}{n\cdot p \,\omega^{\prime}}
+ \frac{\pi^{2}}{6} - 1
\Big)\,
\phi_{D,}^{+}(\omega^{\prime},\mu)
\nonumber \\
&& + \,
\Big(2\ln\frac{\mu^{2}}{n\cdot p\,\omega^{\prime}} + 3 \Big)
\,\omega^{\prime}
\int^{\infty}_{\omega^{\prime}} d\omega
\ln\frac{\omega-\omega^{\prime}}{\omega^{\prime}}
\frac{d}{d\omega} \frac{\phi_{D}^{+}(\omega,\mu)}{\omega}
\nonumber \\
&& - \,2\ln\frac{\mu^{2}}{n\cdot p\,\omega^{\prime}}
\int^{\omega^{\prime}}_{0} d\omega
\ln\frac{\omega^{\prime}-\omega}{\omega^{\prime}}
\frac{d}{d\omega} \phi_{D}^{+}(\omega,\mu)
\nonumber\\
&& +\,\int^{\omega^{\prime}}_{0}d\omega
\ln^{2}\frac{\omega^{\prime}-\omega}{\omega^{\prime}}
\frac{d}{d\omega}
\Big[
\frac{\omega^{\prime}}{\omega} \phi_{D}^{+}(\omega,\mu)
+ \phi_{D}^{+}(\omega,\mu)
\Big]
\bigg\}.
\label{eq:phieff}
\end{eqnarray}

\subsection{The local sub-leading power contribution and higher-twist contribution}
In this section, we will compute the higher-twist contribution and the local sub-leading power contribution of the radiative $D$-meson decay. The local sub-leading power contribution in this procedure is given by evaluating the second diagram and the local term of the first diagram in figure \ref{treelevelfig}, which is the same as \cite{Beneke:2011nf}  by just changing bottom quark to charm quark according to the symmetry of heavy quarks
\begin{eqnarray}
F_{V, \rm NLP}^{\rm {LC}}(n \cdot p)
= -\hat{F}_{A, \rm  NLP}^{\rm {LC}}(n \cdot p)
= \frac{Q_d \, f_D \, m_D}{(n \cdot p)^2}
+ \frac{Q_c \, f_D \, m_D}{n \cdot p \, m_c}  \,,
\label{subleading power local contribution}
\end{eqnarray}
where the first term and second term correspond to the photon emission from the down quark and charm quark, respectively.

The higher-twist contribution is from the non-local term of propagator in the first diagram in figure \ref{treelevelfig}. The hadronic tensor in the frame work of HQET reads
\begin{eqnarray}
T_{\mu\nu}^{(u)}(p,q) &=&
-iQ_q \int d^4x\,e^{ipx}\langle 0 | T\{ \bar d(x) \gamma_\mu d(x),\,
\bar{d}(0)\gamma_\nu(1-\gamma_5)h_v(0)\}| D(p+q)\rangle + \ldots \qquad ,
\label{eq:HQE}
\end{eqnarray}
where the following tree level  matching of the heavy-to-light currents is used
\begin{equation}\label{effective quark field}
\bar{d} \gamma_{\mu} c=\bar{d}\gamma_{\mu}h_v+\ldots\,.
\end{equation}
We will consider the contribution from two-particle twist-4 and three-particle LCDA of $D$-meson. In the calculation of the contribution from three-particle LCDA of $D$-meson,  the light-cone expansion of the quark propagator \cite{Braum:1989ii} in (\ref{eq:HQE}) is required
\begin{eqnarray}
\wick{1}{<1 d(x) >1 {\overline{d}}(0)} &=&
\frac{i}{2\pi^2} \frac{\slashed{x}}{x^4}
-\frac{1}{8\pi^2x^2}
\int_0^1du \bigg\{ix^\rho g\widetilde G_{\rho\sigma}(ux)\gamma^\sigma\gamma_5
+ (2u-1) x^\rho gG_{\rho\sigma}(ux)\gamma^\sigma\biggr\}  +\ldots\,
\nonumber\\[-0.1cm]
\\[-0.8cm]
\nonumber
\end{eqnarray}
Inserting the above propagator into the correlation function (\ref{eq:HQE}), we  can obtain the factorization formula of the higher-twist contribution. At tree level, the factorization formula can be further simplified by taking advantage of the QCD equation of motion to relating the two-particle and three particle LCDAs. Finally, we arrive at the higher-twist contributions
\begin{equation}\label{3p}
\begin{split}
 F_{V,\rm NLP}^{\rm{HT}}(n \cdot p)&=\hat{F}_{A,\rm NLP}^{\rm{HT}}(n \cdot p)=
 -\frac{2 Q_d f_D m_D}{(n\cdot p)^2}\left [\frac{2(\lambda_E^2+2\lambda_H^2)}{6\bar{\Lambda}^2+2\lambda_E^2+\lambda_H^2}+\frac{1}{2}\right ],
\end{split}
\end{equation}
where $\lambda_E^2$ and $\lambda_H^2$ denote the higher-twist matrix elements which is defined by
\begin{eqnarray}
\lefteqn{\langle 0| \bar q(0) g_sG_{\mu\nu}(0)\Gamma h_v(0)|D(p+q)\rangle =}
\nonumber\\&=&
-\frac{i}{6} F_D \lambda^2_H {\rm Tr}\Big[\gamma_5\Gamma P_+ \sigma_{\mu\nu}\Big]
-\frac{1}{6} F_D\Big( \lambda^2_H- \lambda^2_E\Big)
  {\rm Tr}\Big[\gamma_5\Gamma P_+(v_\mu\gamma_\nu-v_\nu\gamma_\mu)\Big]\,,\quad
\label{def:lambdaEH}
\end{eqnarray}
and $\bar{\Lambda}=m_D-m_c$. This expression is the same as the first term in \cite{Beneke:2018wjp}.

Collecting the results of (\ref{NLL 2-particle contribution to form factors}), (\ref{subleading power local contribution}) and (\ref{3p}) together, we get the form factors of $D$-meson decay including the NLP corrections
\begin{eqnarray}
F_{V}(n \cdot p) = F_{V,\rm LP}^{\rm NLL}(n \cdot p) +F_{V,\rm NLP}^{\rm soft}(n \cdot p)  + F_{V,\rm  NLP}^{\rm LC}(n \cdot p)+F_{V,\rm NLP}^{\rm HT}(n \cdot p) \,,
\label{final expression of FV}  \\
\hat{F}_{A}(n \cdot p) = \hat{F}_{A, \rm LP}^{\rm NLL}(n \cdot p) + \hat{F}_{A,\rm NLP}^{\rm soft}(n \cdot p)
+ \hat{F}_{A,\rm NLP}^{\rm LC}(n \cdot p) +\hat{F}_{A,\rm NLP}^{\rm HT}(n \cdot p)\,.
\label{final expression of FAhat}
\end{eqnarray}

\subsection{Power counting}

Following \cite{Beneke:2000aaaa}, the power counting scheme relies on the behaviour of the $D$-meson DA
\begin{eqnarray}
\phi_D^{+}(\omega, \mu_0)  \sim \left\{
\begin{array}{l}
{1 / \Lambda}  \,; \qquad
\omega \sim \Lambda \vspace{0.4 cm} \\
0 \,; \qquad  \hspace{0.5 cm} \omega \gg \Lambda
\end{array}
 \hspace{0.5 cm} \right. ,
\end{eqnarray}
implying that the power counting of the inverse moment is $\lambda_D(\mu_0)\sim\Lambda$ with $\omega\sim\Lambda$. However, the scaling of the inverse moment should be $\lambda_D(\mu_0)\sim \Lambda^2/m_c$ with $\lambda_D(\mu_0)\leq100\, \rm MeV$ in the heavy quark limit, and this will lead to
\begin{eqnarray}
F_{V}^{\rm LP} \sim F_{V}^{\rm soft} \sim
\left ( {m_c \over \Lambda} \right )^{1/2} \,, \qquad
{\rm for} \hspace{0.5 cm} \lambda_D(\mu_0) \sim \Lambda^2/m_c \,,
\end{eqnarray}
this region will be shown in our numerical analysis of $\lambda_D$ dependence. When $\lambda_D(\mu_0)\geq200\,\rm MeV$, the power counting scheme will become
\begin{eqnarray}
F_{V, \rm 2P}^{\rm{LP}} \sim \left ( {\Lambda \over m_c} \right )^{1/2} \,, \qquad
F_{ V,\rm 2P}^{\rm{NLP}} \sim \left ( {\Lambda \over m_c} \right )^{3/2} \,, \qquad
{\rm for} \hspace{0.5 cm} \lambda_D(\mu_0) \sim \Lambda \,,
\end{eqnarray}
which is consistent with typical power counting rules.

\begin{figure}
  \centering
  \includegraphics[width=0.7\textwidth]{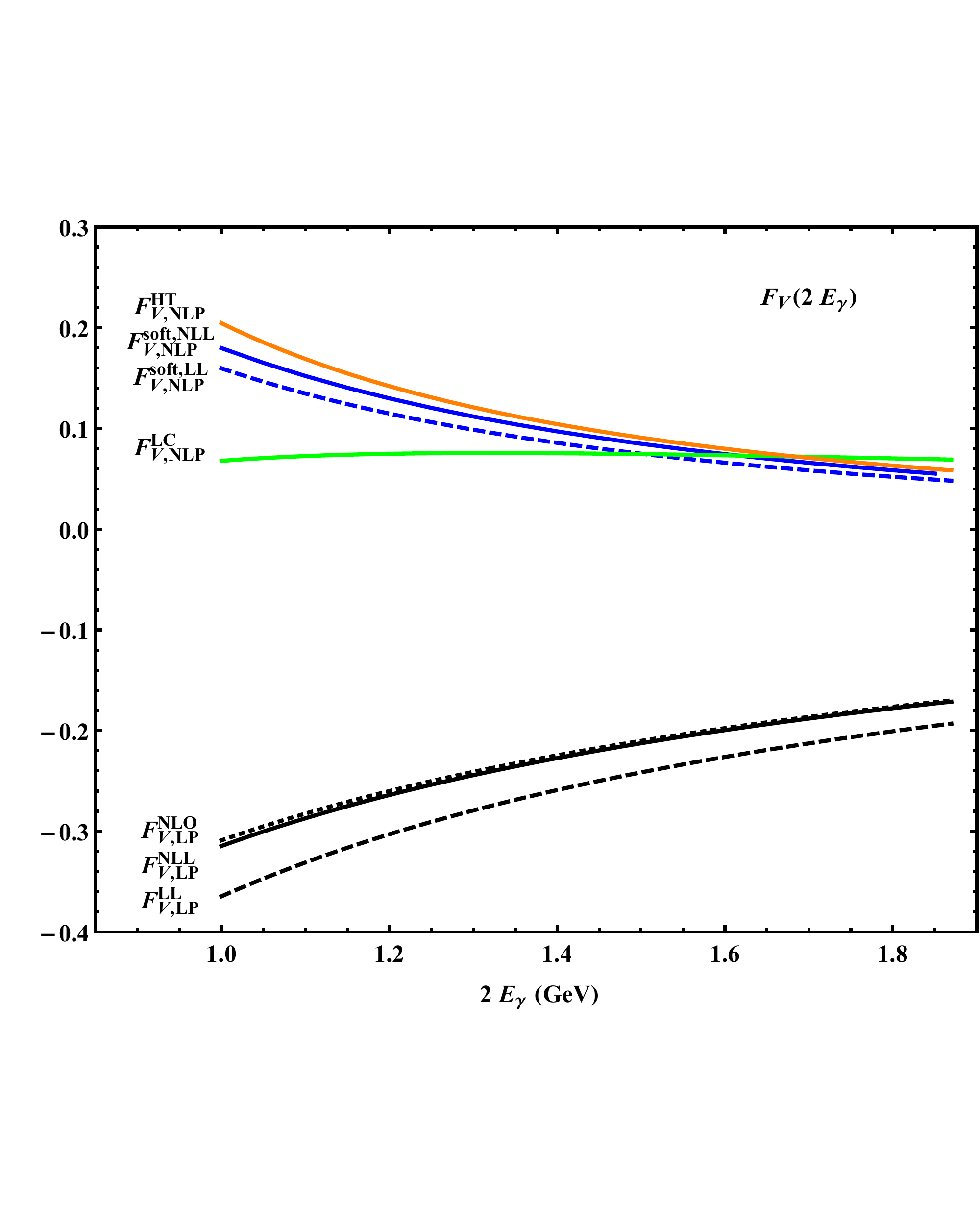}
  \caption{The photon-energy dependence of various contributions to the form factor $F_V(2E_{\gamma})$, with the   exponential model of $\phi_D^{+}(\omega, \mu_0)$ and the inverse moment $\lambda_D(\mu_0)$ =354 MeV.}
  \label{FF1}
\end{figure}

\begin{table}[htb]
\centering{}
\begin{tabular}{c|c|c|c}
\hline\hline
Parameter          & DATA                                & Parameter                  & DATA \tabularnewline
\hline
$m_D $             & $1.86965\pm0.05\, \rm GeV $         &  $n\cdot p\,\omega_s$       & $(1.5\pm0.2)\,\rm GeV^2  $    \tabularnewline
$\tau_D$           & $(1.040\pm0.007)\times10^{-12}\,s$  &  $\mu_{h1}$                 & $1.288\, \rm GeV         $    \tabularnewline
$m_c$              & $1.288\pm0.020\, \rm GeV$           &  $\mu_{h2}$                 & $1.288\, \rm GeV         $    \tabularnewline
$m_d$              & $4.71\pm0.09\, \rm MeV $            &  $\mu_0$                    & $1\, \rm GeV             $    \tabularnewline
$|V_{cd}|$         & 0.221$\pm$0.004                     &  $\mu$                      & $(1.2\pm0.2)\,\rm GeV      $    \tabularnewline
$\lambda_D(\mu_0)$ & $0.354_{-0.03}^{+0.038}\,\rm  GeV$  &  $f_D$                      & $212.0\pm0.7\, \rm MeV   $    \tabularnewline
$\sigma_1(\mu_0)$  & $1.5\pm1$                           &  $\lambda_E^2/\lambda_H^2$  & $0.5\pm0.1\,\rm         $
\tabularnewline
$\sigma_2(\mu_0)$  & $3\pm2$                             &  $2\lambda_E^2+\lambda_H^2$ & $0.25\pm0.15\,\rm GeV^2  $    \tabularnewline
$n\cdot p\,\omega_M$& $(1.25\pm 0.25)\,\rm GeV^2$        & $\bar{\Lambda}$             & $0.58\, \rm GeV          $    \tabularnewline

\hline
\end{tabular}\caption{ Various parameters employed in our calculation, where the hadron masses and life time are from \cite{PDG:2020} and the others are from \cite{Wang:2016qii}. }
\centering{}
\label{values}
\end{table}

\section{Numerical analysis}

We consider two models of the two-particle $D$-meson DA $\phi_D^+(\omega,\mu_0)$ \cite{Wang:2016qii} in our calculation
\begin{eqnarray}
&&  \phi_{D,\rm I}^+(\omega,\mu_0) = \frac{\omega}{\omega_0^2} \, e^{-\omega/\omega_0} \,,
\label{the first model of the B-meson DA} \\
&&  \phi_{D,\rm II}^+(\omega,\mu_0)= \frac{1}{4 \pi \,\omega_0} \, {k \over k^2+1} \,
\left[ {1 \over k^2+1} - \frac{2 (\sigma_{1}(\mu_0) -1)}{\pi^2}  \, \ln k \right ] \,,
\label{the second model of the B-meson DA}
\end{eqnarray}
where $\omega_0=\lambda_D(\mu_0)$, and $k= \omega/(1 \,\, \rm GeV)$. $\phi_{D,\rm I}^+(\omega,\mu_0)$ and   $\phi_{D,\rm II}^+(\omega,\mu_0)$ are based upon the Grozin-Neubert parametrization (left panel) and Braun-Ivanov-Korchemsky parametrization (right panel) respectively. The value of $\lambda_D(\mu_0)$ is taken from the lattice simulations \cite{Aoki:2016frl}.
From the one loop evolution equation of $\phi_D^+(\omega,\mu)$ \cite{Lange:2003aaa,Braun:2004aaa}, we derive the scale-dependence of the inverse moment
\begin{eqnarray}
\frac{\lambda_D(\mu_0)}{\lambda_D(\mu)} &=&
1 + {\alpha_s(\mu_0) \, C_F \over 4 \, \pi} \, \ln {\mu \over \mu_0} \,
\left [2 - 2\, \ln {\mu \over \mu_0} - 4 \, \sigma_{1}(\mu_0) \right ] + {\cal O}(\alpha_s^2)\,.
\label{lambdab evolution}
\end{eqnarray}
In the above, $\sigma_1(\mu_0)$ ($\sigma_2(\mu_0)$ in (\ref{QCDF for B to gamma FFs})) is inverse-logarithmic moment. The definition of the inverse-logarithmic moment \cite{Beneke:2011nf} is
\begin{equation}\label{inv-log}
  \sigma_n(\mu)=\lambda_D(\mu)\int_{0}^{\infty}\frac{d\omega}{\omega}{\rm ln}^n\frac{\mu_0}{\omega}\phi_{D}^+(\omega,\mu),
\end{equation}
where $\mu_0$ is fixed at $1\, \rm GeV$. One could find the other numerical values of the input parameters in table \ref{values}, the factorization scale interval is $\mu=1.2\pm0.2\,\rm GeV$, and the hard scale ($\mu_{h1}=\mu_{h2}={\cal O}(m_c)$) interval is $m_c/2\sim 2\,m_c$.

We will take $\phi_{D,\rm I}^+$ as the default model in the following analysis. With $\lambda_D(\mu_0)=354\,\rm MeV$ and the kinematic region $n\cdot p\in [ 1 \,{\rm  GeV}, m_D ]$, we evaluate the sub-leading power contributions of $D\to\gamma \,\ell \,\nu$. As is shown in figure \ref{FF1}, the sub-leading power contributions are sizeable. The higher-twist contribution reduce the leading power contribution about $(35\sim 65)\%$, and the soft contribution $F_{V,\rm NLP}^{\rm soft}$ leads to a $(30\sim 60)\%$ reduction in the leading power contribution $F_{V,\rm LP}^{\rm NLL}$. The local sub-leading power contribution is insensitive to the photon energy, and the correction to the LP form factor is about $(20\sim 40)\%$. Comparing the soft vector form factor $F_{V,\rm NLP}^{\rm soft,LL}$  with $F_{V,\rm NLP}^{\rm soft,NLL}$, we find the perturbative QCD correction gives rise to about $12.5\%$ shift compared to the leading logarithm (LL) contribution. From the above discussion, we conclude that the leading power contribution is mainly corrected by the soft and the higher-twist contributions at low photon energy, and the local sub-leading correction to the LP form factor will be enhanced at high photon energy.

\begin{figure}
  \centering
  \includegraphics[width=0.45\textwidth]{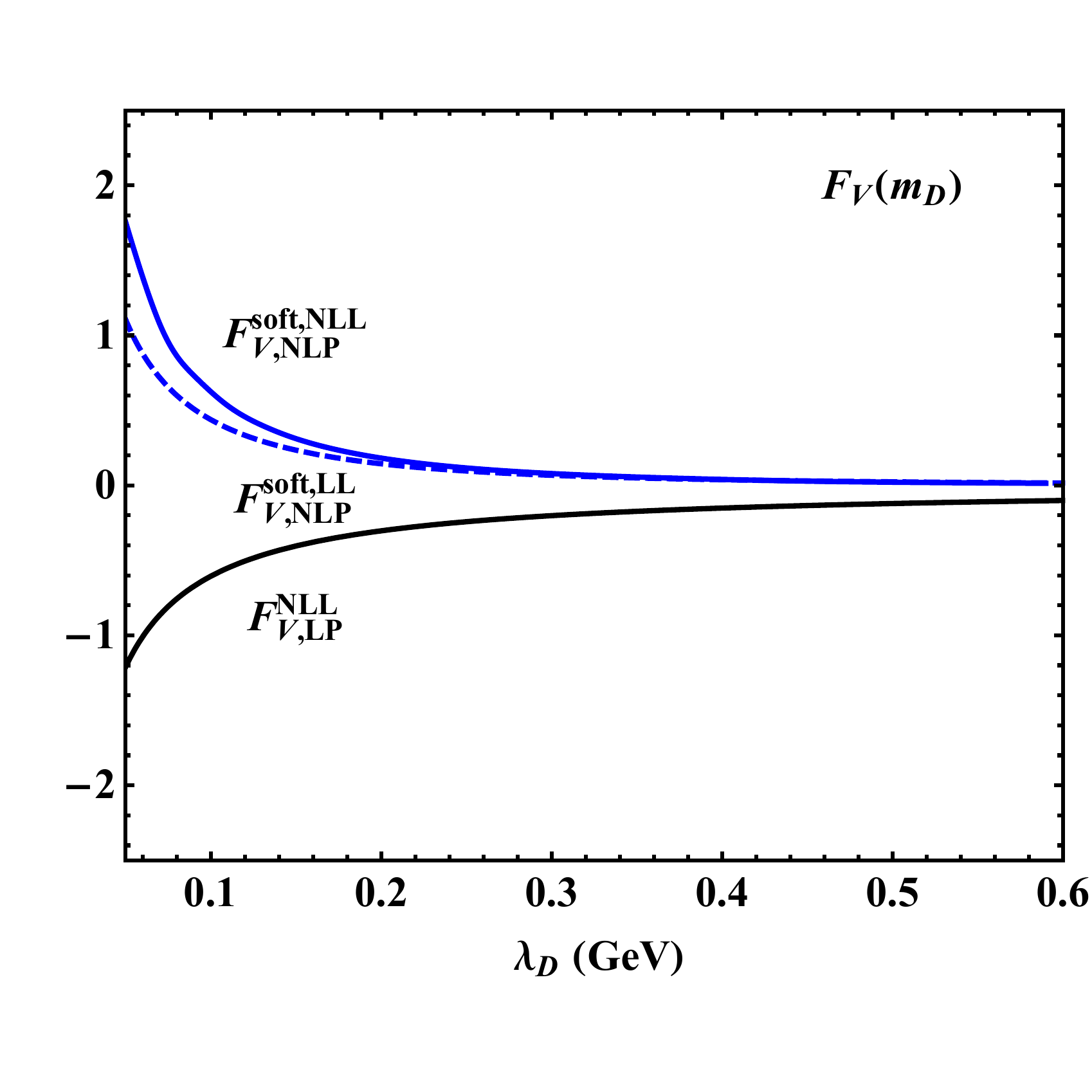}
  \includegraphics[width=0.45\textwidth]{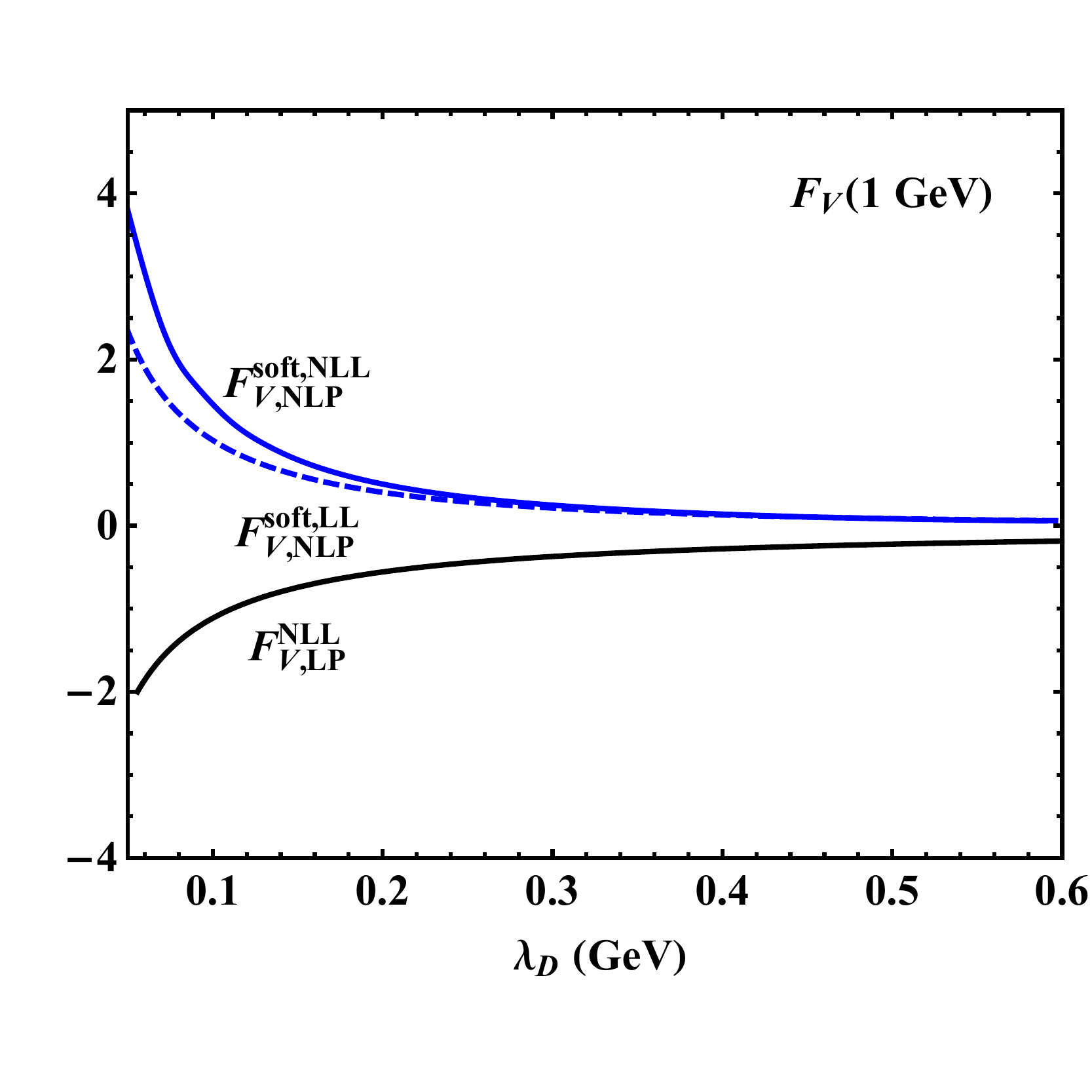}
  \caption{Dependence of the leading and subleading power two-particle contributions to the
form factor $F_V(n\cdot p)$ on the inverse moment $\lambda_D(\mu_0)$ at zero momentum transfer (left panel) and at $n\cdot p$=1 GeV (right panel).}\label{FF2}
\end{figure}

Now we investigate the $\lambda_D(\mu_0)$ dependence of the sub-leading power contributions.
As is shown in figure \ref{FF2}, the form factor $F_{V,\rm NLP}^{\rm soft,NLL}$ decreases rapidly when $\lambda_D\leq 0.15 \,\rm GeV$.
With $\lambda_D=0.1\, \rm GeV$, the form factor $F_{V,\rm NLP}^{\rm soft,NLL}$ can decrease the leading power contribution about $100\%$ at $n\cdot p=m_D$ and $130\%$ at $n\cdot p=1\, \rm GeV$.
The NLL resummation will give a sizable correction to $F_{V,\rm NLP}^{\rm soft,LL}$, both of ${\cal O}(50\%)$ with $\lambda_D(\mu_0)=0.1\,\rm GeV$ and $\lambda_D(\mu_0)=m_D$.
The higher-twist correction to the form factor $F_V$ at $\lambda_D=0.1\, \rm GeV$ is about $10\%$ at $E_{\gamma}=m_D$ and $18\%$ at $E_{\gamma}=1\, \rm GeV$, and this result could be explained by the analytical expression (\ref{3p}). As the results are insensitive to the inverse moment of D meson LCDA, the higher-twist and local sub-leading power corrections to the leading power form factors will be enhanced with the growing inverse moment. In short, the next-to-leading power contributions are large with small $\lambda_D(\mu_0)$, and the higher-twist and local sub-leading power corrections among them are insensitive to the inverse moment.

\begin{figure}
  \centering
  \includegraphics[width=0.45\textwidth]{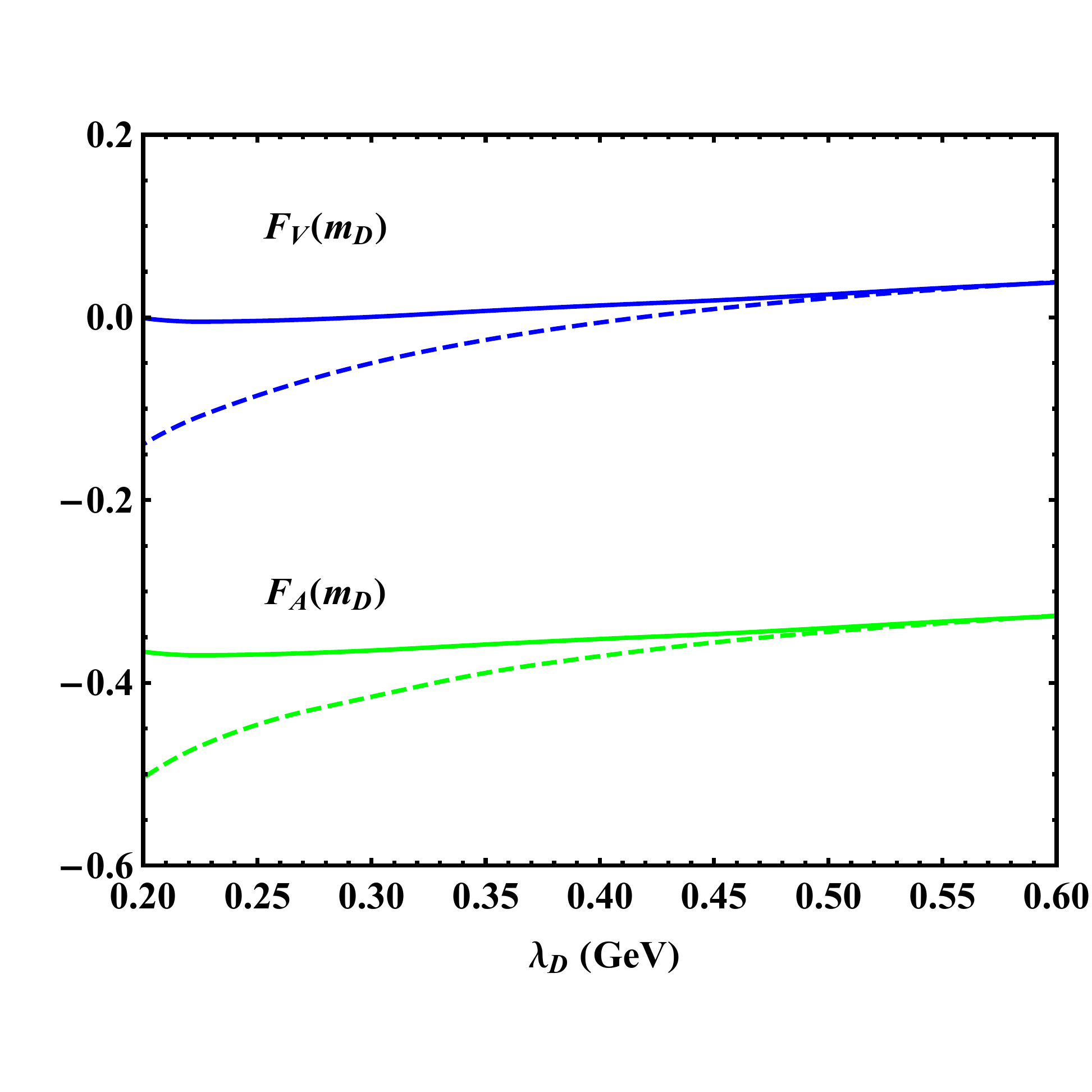}
  \includegraphics[width=0.45\textwidth]{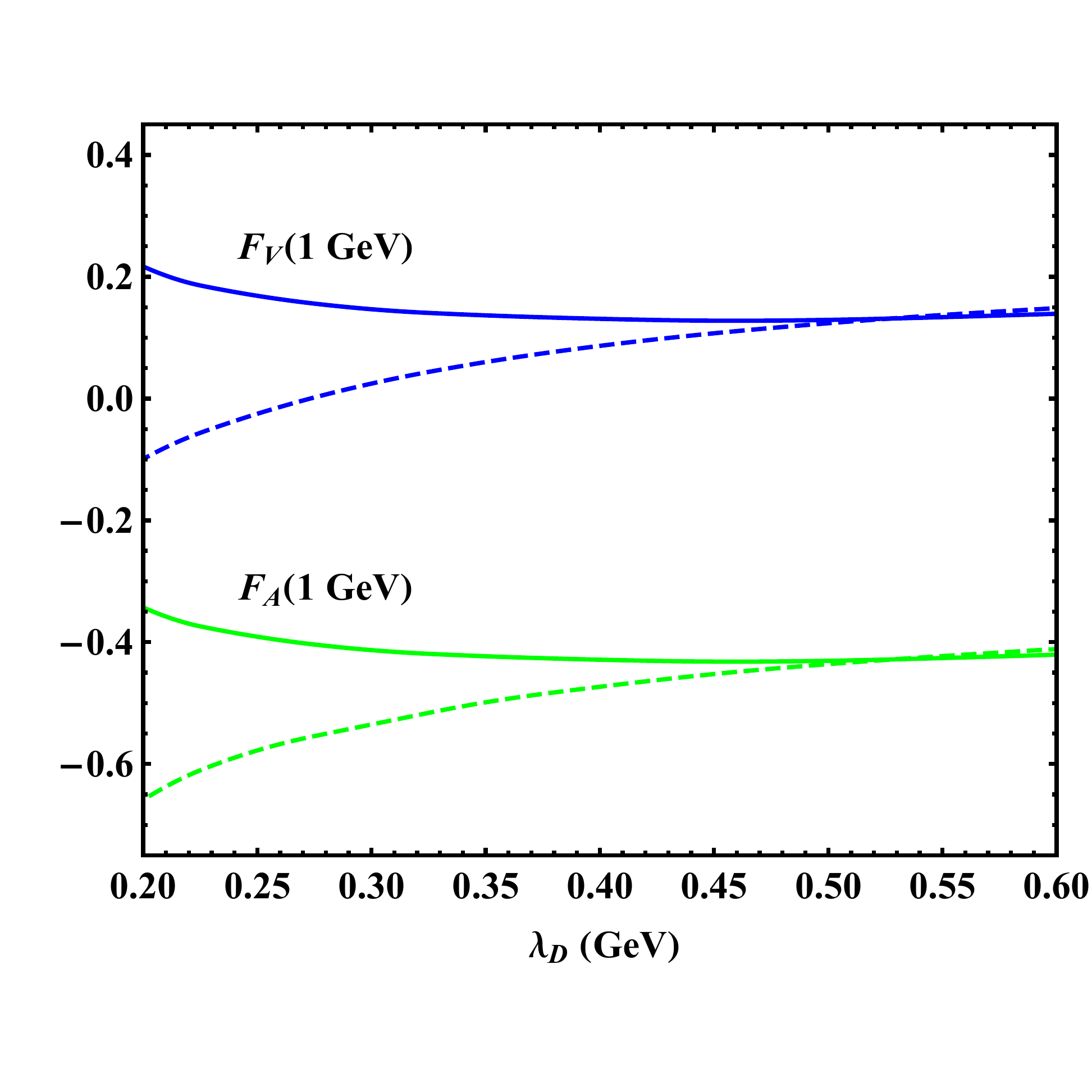}
  \caption{Dependence of the form factors on specific model for $D$-meson DA at $n\cdot p=m_D$ and at $n\cdot p=1\, GeV$. The solid and dashed blue (green) curve indicate the predictions of $F_V(F_A)$ from model $\phi_{D,\rm I}^+$ and model $\phi_{D,\rm II}^+$, respectively.}\label{FF3}
\end{figure}

\begin{figure}
  \centering
  \includegraphics[width=0.5\textwidth]{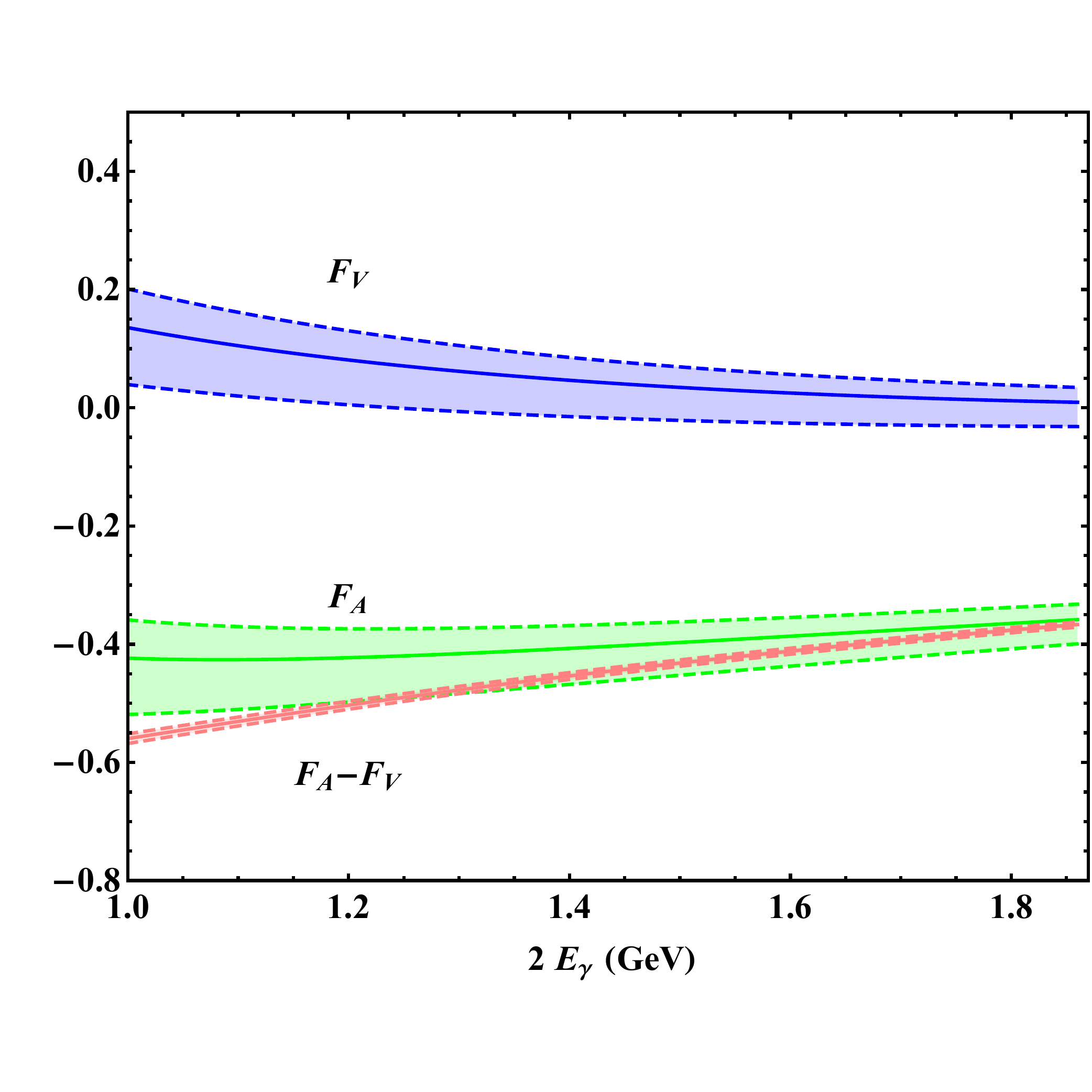}
  \caption{The photon-energy dependence of the form factors $F_V(2\,E_{\gamma})$ and $F_A(2\,E_{\gamma})$ as well as their difference with $\lambda_D(\mu_0)$=354\, MeV. 
  }  \label{FF5}
\end{figure}

We have discussed the $\lambda_D(\mu_0)$ dependence of the form factors $F_V$ and $F_A$ in detail, and now we will focus on the dependence on the $D$-meson LCDA models and the photon energy.
From figure \ref{FF3}, we find that both $F_V$ and $F_{\rm A}$ are insensitive to the models for a large value of inverse moment. This could be easily concluded from figure \ref{FF2}, and the discrepancies of the form factor predictions from different models will be enhanced at $n\cdot p=1\,\rm GeV$.
The photon energy dependence of the $D\to\gamma$ form factors $F_V$, $F_A$, and their difference $F_A-F_V$ are shown in figure \ref{FF5}.
In our calculation, the uncertainties arise from the errors of $\mu$, $\lambda_D(\mu_0)$, $\sigma_1(\mu_0)$ and $\sigma_2(\mu_0)$, and different models of $D$-meson $\phi_D^+(\omega,\mu_0)$. It is easy to find that the soft contribution is sensitive to the shape of $\phi_D^+(\omega,\mu_0)$ at small $\omega$ from the analytical expression of $F_{V,\rm NLP}^{\rm soft,NLL}$.
As the soft and higher-twist contributions are symmetry conserved, the symmetry breaking effect originates from the local sub-leading correction
    \begin{eqnarray}
    F_{ A}(n \cdot p)-F_{V}(n \cdot p)={2 \, f_D \over n \cdot p} \,
    \left [Q_{\ell} - {Q_d \, m_D \over n \cdot p} - {Q_c \, m_D \over m_c} \right ]
    + {\cal O}(\alpha_s) \,.
    \end{eqnarray}
We should note that the local sub-leading power correction only depends on the decay constant $f_D$, and this can explain why the uncertainty of $F_{A}-F_{V}$ is so small.

\begin{figure}
  \centering
  \includegraphics[width=0.45\textwidth]{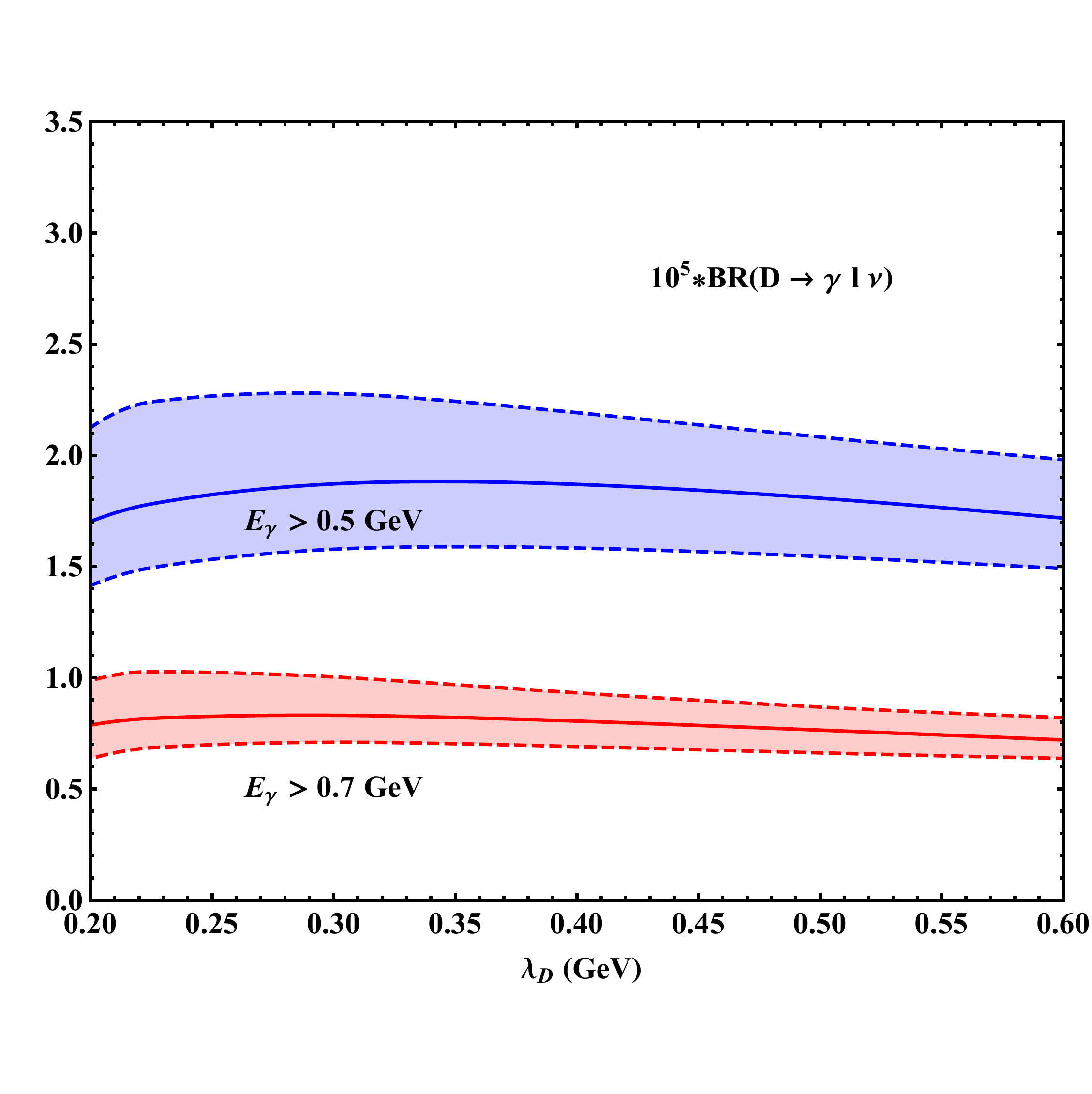}
  \includegraphics[width=0.45\textwidth]{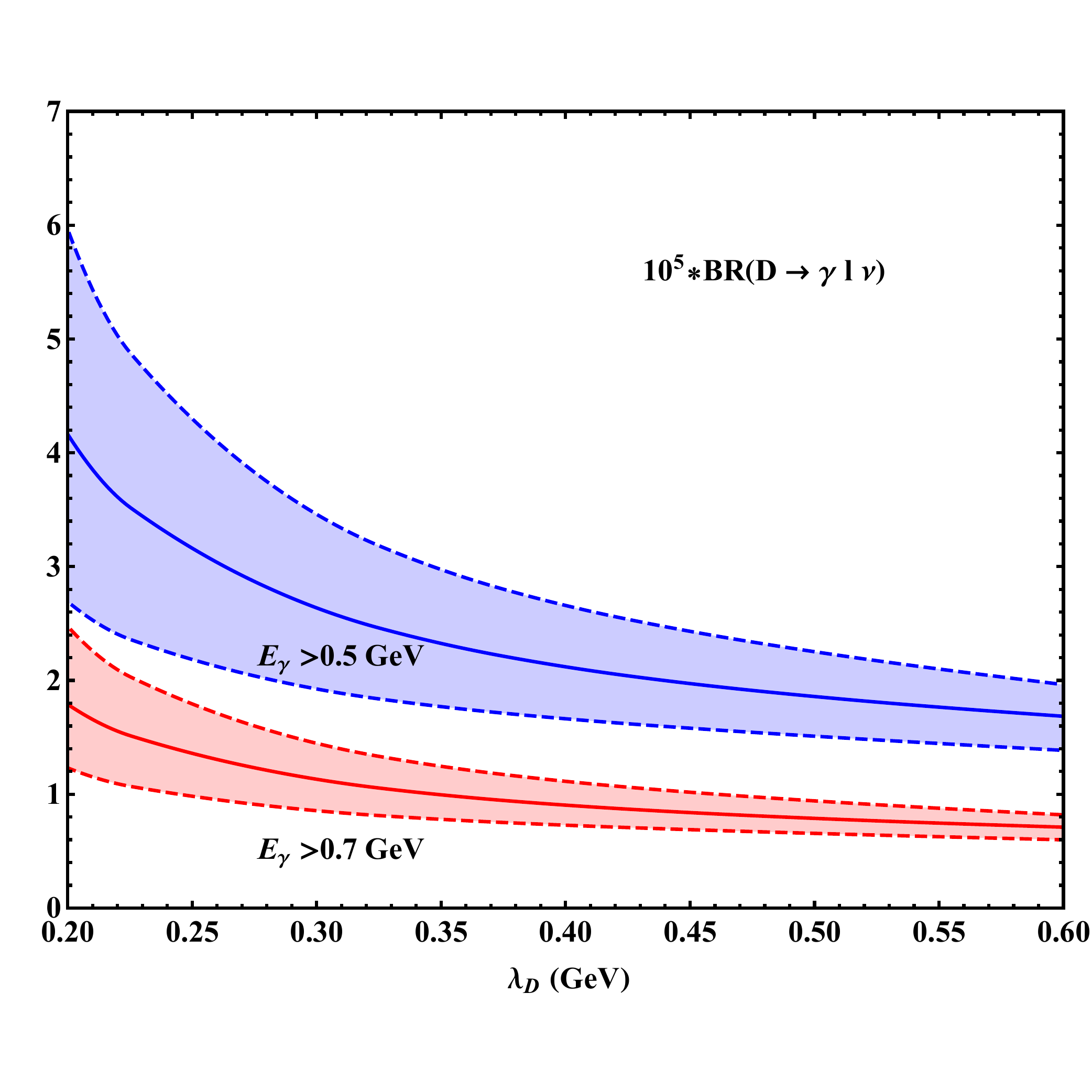}
  \caption{The blue band shows the inverse-moment $\lambda_D(\mu_0)$ dependence of the partial branching fractions of ${\cal BR}(D\rightarrow\gamma l\nu,E_{\gamma}\geq E_{cut})$ for $E_{\rm cut}$= 0.5 GeV with the model $\phi_{D,\rm I}^+(\omega,\mu_0)$ based upon the Grozin-Neubert parametrization (left panel) and with the model $\phi_{D,\rm II}^+(\omega,\mu_0)$ based upon the Braun-Ivanov-Korchemsky parametrization (right panel). The red  band shows the inverse-moment $\lambda_D(\mu_0)$ dependence of the partial branching fractions for $E_{\rm cut}$= 0.7 GeV with the model $\phi_{D,\rm I}^+(\omega,\mu_0)$ (left panel) and with the model $\phi_{D,\rm II}^+(\omega,\mu_0)$ (right panel).}\label{FF6}
\end{figure}

\begin{figure}
  \centering
  \includegraphics[width=0.5\textwidth]{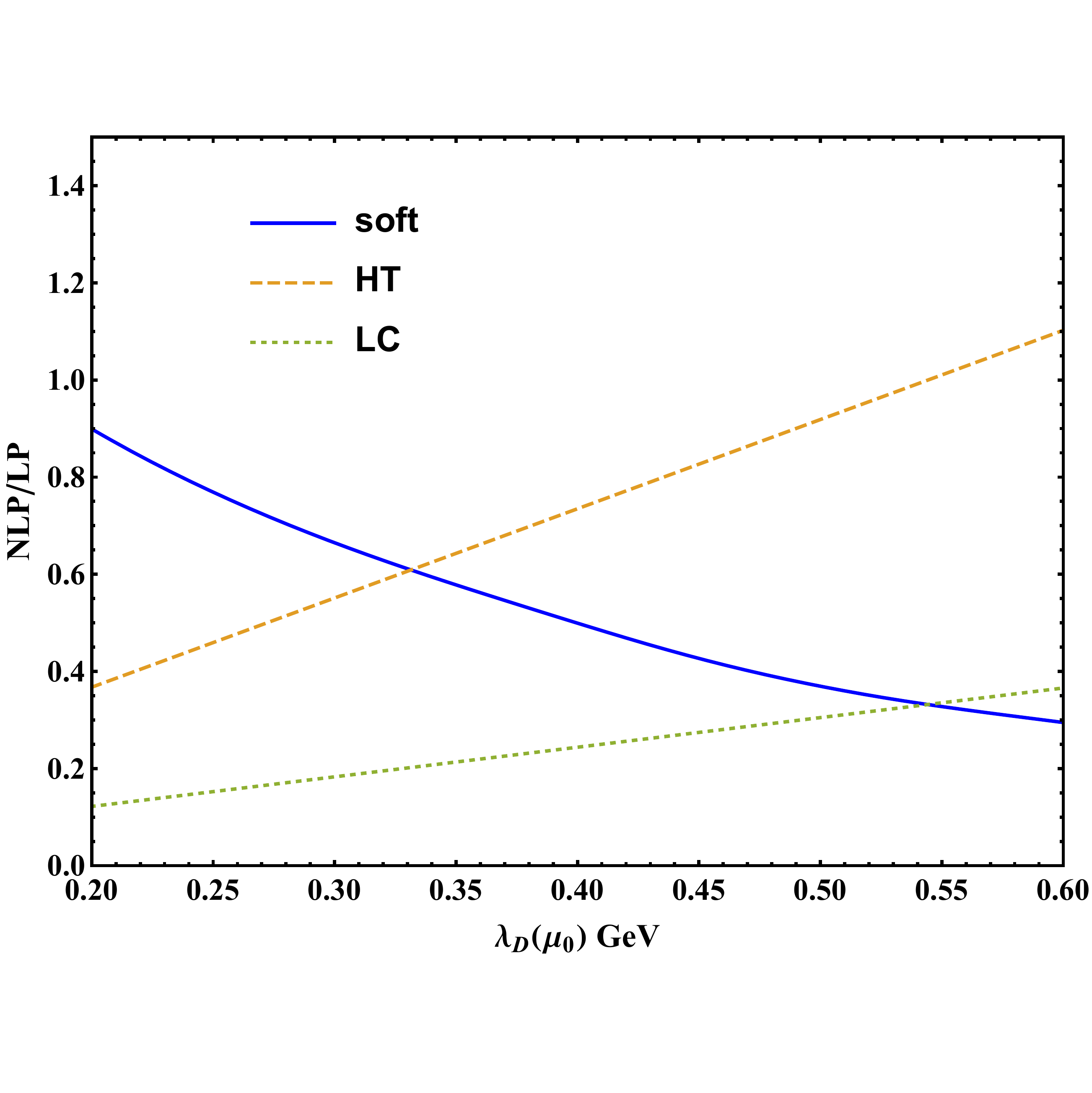}
  \caption{Dependence of the NLP corrections to the LP vector form factor from the $D$-meson DA model $\phi_{D,\rm I}^+(\omega,\mu_0)$ on the inverse moment with $n\cdot p=1\,\rm GeV$. }\label{ratios}
\end{figure}

We consider the theory constraint on $\lambda_D(\mu_0)$ now. Since we have chosen the power counting scheme $n\cdot p=2\,E_{\gamma}\sim{\cal O}(m_D)$ in our calculation of the factorization formula, we should write the integrated decay rate as
\begin{eqnarray}
\Delta {\cal BR}(E_{\rm cut}) = \tau_{D} \, \int_{E_{\rm cut}}^{m_D/2} \, d \, E_{\rm \gamma} \,\,
\frac{d \, \Gamma}{ d \, E_{\rm \gamma}} \left ( D \to \gamma \ell \nu \right ) \,.
\end{eqnarray}
From the BES-III experiment, we have known the upper limit on the branching ratio $\Delta {\cal BR}(E_{\gamma}\geq10\,\rm MeV)<3\times10^{-5}$ with the photon energy larger than $10\, \rm MeV$. As the photon emission off the charged particle is not a soft photon, we can choose $E_{\rm cut}=0.5\,\rm GeV$. One can see from the left panel of figure \ref{FF6} that there is no bound on $\lambda_D(\mu_0)$ for the Grozin- Nuebert mode after including the soft and higher twist contributions. With $\lambda_D(\mu_0)=354\,\rm MeV$, our prediction of the branching fraction is $(1.88_{-0.29}^{+0.36})\times10^{-5}$. The right panel of figure \ref{FF6} shows that branching ratio of DA model $\phi_{D,\rm II}^+(\omega,\mu_0)$ is large at small $\lambda_D(\mu_0)$, and the experimental result yields a bound $\lambda_D(\mu_0)>270\,\rm MeV$. To understand this result, we should note that the behavior of the form factors in the Braun-Ivanov-Korchemsky model is sensitive to the inverse moment of $D$-meson LCDA at small $\lambda_D(\mu_0)$. The prediction of the branching fraction from this model is $(2.31_{-0.54}^{+0.65})\times10^{-5}$. It is difficult to extract the inverse moment of $D$-meson LCDA in the $D\to\gamma\,\nu\,\ell$ decays as the inverse moment dependence of the NLP corrections is complicated.
As is shown in figure \ref{ratios}, the higher-twist and local sub-leading power corrections are enhanced with the growing inverse moment, and the large correction makes it hard to extract the inverse moment.

We have known theoretically that the power corrections of $1/m_c$ will be significant, and now we will estimate the corrections in practice. Although it is less effective to study the radiative leptonic $D$-meson decay by the factorization approach, we can still deepen our understanding of the factorization approach through the $D\to\gamma\,l\,\nu$ process.
If we fix the photon energy at 0.5 GeV, and adopt the $D$-meson LCDA as $\phi_{D,\rm I}^+(\omega,\mu_0)$, the NLP correction to the LP form factors is about $143\%$, and the correction is about $120\%$ when LCDA model $\phi_{D,\rm II}^+(\omega,\mu_0)$ is adopted.
The above results indicate that the power suppressed contributions play important roles in radiative leptonic $D$-meson decays, which is consist with the theoretical analysis.

The theoretical uncertainties from the LCDA parameters are collected in table \ref{unc}, and the errors from these parameters are about $(10\sim20)\%$.
These results suggest that the uncertainties from the inverse-logarithmic moment $\sigma_2(\mu_0)$ are great, while the dependence of the model $\phi_{D,\rm II}^+(\omega,\mu_0)$ on $\sigma_1(\mu_0)$ and $\lambda_D(\mu_0)$ is more remarkable. Comparing results evaluated from the total form factor $F_V$ and $F_A$ with the LP form factor $F_{V,\rm LP}^{\rm NLL}$ and $F_{A,\rm LP}^{\rm NLL}$, it is manifest that sub-leading power contributions yield a correction about ${\cal O}(100\%)$ to the branching fraction.

\begin{table}[htb]
\centering{}
\begin{tabular}{c|c|c|c|c|c}
\hline
\hline
                                     &${\rm BR}\times 10^{6}$& $\lambda_D(\mu_0)[\omega_0]$ & $\mu$ &$\sigma_1(\mu_0)$&$\sigma_2(\mu_0)$       \tabularnewline
\hline
\multirow{2}{*}{$\phi_{D,\rm I}^+$}  & \multirow{2}{*}{18.8} & $+0.0 $    & $+0.74$& $+0.59$        &$+2.01$       \tabularnewline
                                     &                       & $-0.08$    &$-0.87$& $-0.36$        &$-1.56$    \tabularnewline
\hline
\multirow{2}{*}{$\phi_{D,\rm II}^+$} & \multirow{2}{*}{23.1} & $+1.64$    &$+0.53$& $+4.73$        &$+2.41$  \tabularnewline
                                     &                       & $-1.58$     &$-0.9 $& $-3.88$        &$-2.17$      \tabularnewline
\hline
\hline
\multirow{2}{*}{$\rm LP \ results$}  & \multirow{2}{*}{12.5} & $+2.4$     &$+0.0 $& $+0.45$        &$+2.45$ \tabularnewline
                                     &                       & $-2.2$     &$-0.46 $& $-0.68$        &$-2.44$  \tabularnewline
\hline
\end{tabular}\caption{ The branching fraction uncertainties with $\lambda_D(\mu_0)=354\,\rm MeV$ associated with the inverse moment, the factorization scale, and two inverse-logarithmic moments $\sigma_1(\mu_0)$ and $\sigma_2(\mu_0)$. The LP results are evaluated from $F_{V,\rm LP}^{\rm NLL}$ and $F_{A,\rm LP}^{\rm NLL}$, others are evaluated from $F_V$ and $F_A$. }\label{unc}
\end{table}

\begin{table}[htb]
\centering{}
\begin{tabular}{c|c|c|c}
\hline\hline
Method                      & BR                                      & Method         & BR                             \tabularnewline
\hline
Model $\phi_{D,\rm I}^+$    & $(1.88_{-0.29}^{+0.36})\times10^{-5}$   & NRQM\cite{Lu:2002mn}         & $4.6\times10^{-6}$             \tabularnewline
Model $\phi_{D,\rm II}^+$   & $(2.31_{-0.54}^{+0.65})\times10^{-5}$   & pQCD\cite{Korchemsky:1999qb} & $(0.82\pm0.65)\times10^{-4}$   \tabularnewline
LFQM \cite{Geng:2000if}     & $2.5\times10^{-5}$                      & RIQM\cite{Barik:2009zza}     & $3.34\times10^{-5}$            \tabularnewline
QCDF\cite{Yang:2016wtm}     & $1.92\times10^{-5}$                     & BES-III\cite{Ablikim:2017twd}& $<3\times10^{-5}$              \tabularnewline
\hline
\end{tabular}\caption{ Results from different methods. }\label{Results Models}
\end{table}

Now we make a comparison with other works. Numerical results of different methods have been collected in table \ref{Results Models}.
Except for the pQCD and RIQM results, we can find results from various methods are consistent with the experimental upper limit. The work of NRQM \cite{Lu:2002mn} is an extension of \cite{Atwood:1994za} by including the diagrams of a photon emission from the heavy quark, the lepton and $W$-boson, which leads to a much smaller result.
In \cite{Geng:2000if}, the LFQM was used to calculate the $D\to\gamma$ form factor, and gave a reliable prediction of the $D$-meson decay.
The ${\cal O}(\Lambda_{\rm QCD}/m_D)$ correction in the factorization method was calculated in \cite{Yang:2014rna}, which is extended in \cite{Yang:2016wtm} by including the soft photon region. Predictions of these two works have been verified by experiment.
We improve the factorization calculation to the NLP corrections, and our predictions of the branching fraction are in agreement with the experimental upper limit. However, as is shown in table \ref{unc}, the sub-leading power contributions are important in this work.
From the above results, we concluded that our results are still bellow the upper limit of the experiment result, and predictions of branching fraction with $E_{\gamma}>0.5\,\rm GeV$ need further verification from the experiment.

\section{Conclusion}

We studied the NLP contributions of the radiative leptonic $D$-meson decay within the framework of factorization. In the study of the $D$-meson decay, both the QCD correction and the power correction are large since the charm quark mass $m_c$ is not large enough. After including the NLP corrections, the theoretical prediction is highly improved. In addition, we provided the analytic expressions of the NLP form factors for $D\to\gamma \,\ell\,\nu$ with the soft contribution, the power suppressed local contribution and the higher-twist contribution included, and the error estimate from the expansion of $m_c$.

By using the model based on the Grozin-Neubert parametrization, the power suppressed correction is dominanted by the soft and higher-twist contributions with $\lambda_D(\mu_0)=354\,\rm MeV$. The experimental upper limit yields a bound $\lambda_D(\mu_0)>270\,\rm MeV$ according to the dependence of branching fractions of DA model $\phi_{D,\rm II}^+$ on the inverse moment of $D$-meson LCDA, but the importance of the power corrections indicates that it is difficult to extract the appropriate inverse moment. Numerically, we found that all the sub-leading power contributions  are significant at $\lambda_D(\mu_0)=354\,\rm MeV$, and the next-to-leading power contributions will lead to $143\%$ in $\phi_{D,\rm I}^+$ and $120\%$ in $\phi_{D,\rm II}^+$ corrections to leading power vector form factors with $E_{\gamma}=0.5\,\rm GeV$.

To summarize, the branching fraction predictions in this work are in agreement with the experimental upper limit, though the NLP corrections are significance. The effects of the power corrections need a further study both in theoretical and experimental, and we hope experiment on $E_{\gamma}>0.5\,\rm GeV$ could be conducted in the future. There exist some other sources of the power correction, such as the power suppressed heavy quark field, the non-local power suppressed terms in the light quark propagator, etc., they will be investigated in future studies.

\subsection*{Acknowledgements}
This work is supported in part by the National Natural Science Foundation of China with Grant No. 11675082 and 11735010, and the Natural Science Foundation of Tianjin with Grant No. 19JCJQJC61100. The author would like to thank Yu-Ming Wang for illuminating discussions.

\appendix

\section{Renormalization group evolution factor}
\label{b}
This expression has been calculated in \cite{Beneke:2011nf}, and the second factor just need us to set the cusp anomalous dimension to zero, details could be found in this reference.
\begin{eqnarray}
U_1(E_\gamma,\mu_h,\mu) &=&
\exp\left(\,\int_{\alpha_s(\mu_h)}^{\alpha_s(\mu)} d\alpha_s\,
\left[
\frac{\gamma(\alpha_s)}{\beta(\alpha_s)} +
\frac{\Gamma_{\rm cusp}(\alpha_s)}{\beta(\alpha_s)}
\left(
\ln\frac{2 E_\gamma}{\mu_h} -
\int_{\alpha_s(\mu_h)}^{\alpha_s}
\frac{d\alpha_s^\prime}{\beta(\alpha_s^\prime)}
\right)
\right]\right)
\nn \\[0.4cm]
&& \hspace*{-2.2cm}
=\,\exp\left(\,
-\frac{\Gamma_0}{4\beta_0^2} \left(
\frac{4\pi}{\alpha_s(\mu_h)}\left[\ln r-1+\frac{1}{r}\right]
-\frac{\beta_1}{2 \beta_0} \,\ln^2 r
+\left(\frac{\Gamma_1}{\Gamma_0}-\frac{\beta_1}{\beta_0}\right)
\left[r-1-\ln r\right]\right)
\right)
\nn \\[0.2cm]
&& \hspace*{-1.7cm}
\times\,\left(\frac{2 E_\gamma}{\mu_h}\right)^
{-\frac{\Gamma_0}{2\beta_0} \ln r}
 r^{-\frac{\gamma_0}{2\beta_0}}
\times \Bigg[1 - \frac{\alpha_s(\mu_h)}{4\pi}\,\frac{\Gamma_0}{4\beta_0^2}
\,\bigg(\frac{\Gamma_2}{2\Gamma_0} \left[1-r\right]^2
+\frac{\beta_2}{2\beta_0} \left[1-r^2+2 \ln r\right]
\nn \\[0.2cm]
&& \hspace*{-0.7cm}
-\,\frac{\Gamma_1\beta_1}{2\Gamma_0\beta_0}
\left[3-4 r+r^2+2 r \ln r\right]
+\frac{\beta_1^2}{2\beta_0^2} \left[1-r\right]\left[1-r-2\ln r\right]
\bigg)
\nn \\[0.2cm]
&& \hspace*{-0.7cm}
+\,\frac{\alpha_s(\mu_h)}{4\pi}\left(
\ln\frac{2E_\gamma}{\mu_h}
\left(\frac{\Gamma_1}{2\beta_0}-\frac{\Gamma_0\beta_1}{2\beta_0^2}\right)
+\frac{\gamma_1}{2\beta_0}-\frac{\gamma_0\beta_1}{2\beta_0^2}\right)
\left[1-r\right] + {\cal O}(\alpha_s^2)\Bigg]
\label{U1}
\end{eqnarray}

\section{Spectral representations}
\label{appendix a}
We collect this dispersion representation of various convolution integral from \cite{Wang:2016qii}.
\begin{eqnarray}
&& {1 \over \pi} \, {\rm Im}_{\omega^{\prime}} \, \int_0^{\infty} \,  \,
\, \frac{d \omega}{\omega-\omega^{\prime}-i0} \,
\ln^2{\mu^2 \over n \cdot p \, (\omega-\omega^{\prime})} \,\, \phi_D^{+}(\omega, \mu) \nonumber \\
&& = \int_0^{\infty} \, d \omega \,
\left [ {2 \, \theta(\omega^{\prime}-\omega) \over \omega - \omega^{\prime}}  \,
\ln {\mu^2 \over n \cdot p \, (\omega^{\prime} - \omega)}   \right ]_{\oplus}
\,\, \phi_D^{+}(\omega, \mu)
+ \left [ \ln {\mu^2 \over n \cdot p \,  \omega^{\prime} }
- {\pi^2 \over 3} \right ] \phi_D^{+}(\omega^{\prime}, \mu) \,, \\
&& {1 \over \pi} \, {\rm Im}_{\omega^{\prime}} \, \int_0^{\infty} \,  \,
\, \frac{d \omega}{\omega-\omega^{\prime}-i0} \,\, { \omega^{\prime} \over \omega } \,\,
\ln {\omega^{\prime} - \omega \over \omega^{\prime}} \,\,
\ln {\mu^2 \over - n \cdot p \, \omega^{\prime}} \,\, \phi_D^{+}(\omega, \mu) \nonumber \\
&& = -{\omega^{\prime} \over 2} \,\,  \bigg \{ \int_0^{\infty} d \omega \,
\ln^2 \bigg|{\omega - \omega^{\prime}  \over \omega^{\prime} } \bigg|  \,\,
{d \over d \omega} \, {\phi_D^{+}(\omega^{\prime}, \mu) \over \omega}  \nonumber \\
&& \hspace{1.5 cm} + \,  \int_{\omega^{\prime}}^{\infty} d \omega \,  \,
\left [ 2 \, \ln{\mu^2 \over n \cdot p \, \omega^{\prime}} \,
\ln {\omega - \omega^{\prime}  \over \omega^{\prime} }  - \pi^2 \right ] \,
{d \over d \omega} \, {\phi_D^{+}(\omega^{\prime}, \mu) \over \omega}  \bigg \} \,, \\
&& {1 \over \pi} \, {\rm Im}_{\omega^{\prime}} \, \int_0^{\infty} \,  \,
\, \frac{d \omega}{\omega-\omega^{\prime}-i0} \,\, { \omega^{\prime} \over \omega } \,\,
\ln {\omega^{\prime} - \omega \over \omega^{\prime}} \,\,
\ln {\mu^2 \over  n \cdot p \, (\omega - \omega^{\prime})} \,\, \phi_D^{+}(\omega, \mu) \nonumber \\
&& = \omega^{\prime} \, \bigg \{ \int_0^{\infty} \, d \omega \,
\left [ { \theta(\omega^{\prime}  - \omega) \over \omega - \omega^{\prime}} \,
\ln { \omega^{\prime} - \omega \over \omega^{\prime}} \right ]_{\oplus} \,
\, {\phi_D^{+}(\omega^{\prime}, \mu) \over \omega}  \nonumber \\
&& \hspace{1.0 cm} + \, {1 \over 2} \, \int_{\omega^{\prime}}^{\infty} \, d \omega \,
\left [ \ln^2 {\mu^2 \over  n \cdot p \, (\omega - \omega^{\prime})}
- \ln^2 {\mu^2 \over  n \cdot p \,  \omega^{\prime}} + {\pi^2 \over 3}  \right ] \,
{d \over d \omega} \, {\phi_D^{+}(\omega^{\prime}, \mu) \over \omega}  \bigg \}  \,, \\
&& {1 \over \pi} \, {\rm Im}_{\omega^{\prime}} \, \int_0^{\infty} \,  \,
\, \frac{d \omega}{\omega-\omega^{\prime}-i0} \,\, { \omega^{\prime} \over \omega } \,\,
\ln {\omega^{\prime} - \omega \over \omega^{\prime}} \,\, \phi_D^{+}(\omega, \mu) \nonumber \\
&& = - \omega^{\prime} \,  \int_{\omega^{\prime} }^{\infty} \, d \omega \,
\ln {\omega - \omega^{\prime} \over \omega^{\prime} } \,\,
{d \over d \omega } \, {\phi_B^{+}(\omega, \mu) \over \omega} \,.
\end{eqnarray}

\end{document}